\documentstyle[epsf,12pt]{article}

\begin{document}

\title{A Soluble Theory of Massless Scalar $QED_{2}$} 

\vspace{1in}

\author{F. T. Brandt$^\dagger$, Ashok Das$^\ddagger$ and
  J. Frenkel$^\dagger$
\\
$^\dagger$Instituto de F\'\i sica,
Universidade de S\~ao Paulo\\
S\~ao Paulo, SP 05315-970, BRAZIL\\
$^\ddagger$Department of Physics and Astronomy,
University of Rochester\\
Rochester, NY 14627, USA}

\date{October   1998}
\maketitle 

\vspace{1in}

\begin{abstract}
In this brief report, we analyze a generalized theory of massless scalar 
$QED_{2}$ and show that, unlike the conventional scalar $QED_{2}$,
it is free from infrared divergence problems. 
The model is exactly soluble and may describe, in an $1+1$
dimensional space-time, noninteracting spin-one tachyons.
 
\end{abstract}

\vfill

\pagebreak

\par 
It is well known that massless particles give rise, in general, to
infrared divergence problems in quantum field theories
\cite{weinberg:book95}.
For example, let us consider the theory of scalar $QED$ 
in $1+1$ dimensions ($QED_2$), described by the Lagrangian density
\begin{equation}
{\cal L} = - {1\over 4} F_{\mu\nu} F^{\mu\nu} + (D_{\mu}\phi)^{*} D^{\mu}\phi
- m^{2}\phi^{*}\phi \label{a1}
\end{equation}
where
\begin{equation}
D_{\mu}\phi = (\partial_{\mu} + ie A_{\mu})\phi\label{a2}
\end{equation} 
\begin{figure}[h!]
\bigskip
   \hspace{.1\textwidth}
  \vbox{\epsfxsize=.5\textwidth
    \centerline{\epsfbox{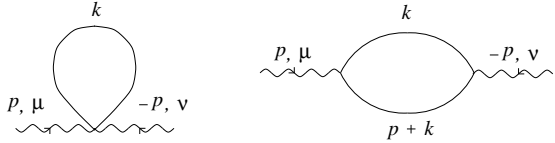}}}
   \label{fig1}
\caption{Diagrams contributing to the photon self-energy at
one-loop order}
  \end{figure}
In this theory, the photon self-energy, at one loop, is given by the
contributions shown in Fig. 1. 
These diagrams are easy to evaluate and give (We would use a gauge invariant
regularization throughout.)
\begin{eqnarray}\label{a3}
\Gamma^{\mu\nu}(p) & = & -i e^{2} \int {d^{2}k\over (2\pi)^{2}}\left[-{2
\eta^{\mu\nu}\over k^{2}-m^{2}+i\epsilon} + {(2k+p)^{\mu}(2k+p)^{\nu}\over
(k^{2}-m^{2}+i\epsilon)((k+p)^{2}-m^{2}+i\epsilon)}\right]\nonumber\\
  &   & \nonumber\\
  & = & {e^{2}\over \pi}\left(\eta^{\mu\nu} - {p^{\mu}p^{\nu}\over p^{2}}
\right)
\left[\frac R 2
\ln{\left(\frac{1+R}{1-R}\right)}-1\right] 
\end{eqnarray}
where
\begin{equation}
R = \sqrt{1-4m^{2}/p^{2}}\label{a4}
\end{equation}
It is clear now that the photon self-energy in Eq. (\ref{a3})
is  ultraviolet finite, but diverges like $\ln{(1/m)}$
as $m\rightarrow 0$. Furthermore, it can be checked that the higher order 
contributions do not lead to an improved infrared behavior of the photon 
self-energy. This is the problem with massless scalar fields in $1+1$  
dimensions and it is well known that even  supersymmetric theories involving 
massless scalar fields suffer from this problem \cite{ferrara}.
In this brief report, 
we present a generalized scalar $QED_{2}$ which is free from infrared 
problems and bring out various other interesting properties associated 
with this system.

Let us consider two flavors of complex scalar fields, $\phi_{a}, a=1,2$, and
the theory described by the Lagrangian density
\begin{equation}
{\cal L} =  - {1\over 4} F_{\mu\nu} F^{\mu\nu} + (D_{\mu}\phi_{a})^{*} 
D^{\mu}\phi_{a} - m^{2}\phi_{a}^{*}\phi_{a} -{ie\over 2}\epsilon^{\mu\nu}
F_{\mu\nu}\phi_{a}^{*}(\sigma_{3})_{ab}\phi_{b} \label{a5},
\end{equation}
where $\epsilon^{\mu\nu}$ is the antisymmetric Levi-Civita  tensor in
$1+1$ dimensions, with $\epsilon^{01}=1$.
This theory differs from the conventional scalar $QED_{2}$ of Eq. (\ref{a1})
(for two flavors) because of the last term in Eq. (\ref{a5}) which has the 
form of a \lq\lq Pauli interaction''. It is again straightforward to calculate
the photon self-energy in this theory where one has to evaluate,
in addition, only the 
contribution coming from the new interaction term.
Instead of the result given in Eq. (\ref{a3}), one finds in this
case that
\begin{equation}
\Gamma^{\mu\nu}(p) = -{2e^{2}\over \pi}\left(\eta^{\mu\nu} - {p^{\mu}p^{\nu}
\over p^{2}}\right)\left[1 +
\frac{2m^2}{p^2} \frac{1}{R}
\ln{\left(\frac{1+R}{1-R}\right)}\right]
\label{a6}
\end{equation}
It is interesting that the second term in the bracket on the right hand side 
of Eq. (\ref{a6}) vanishes when $m\rightarrow 0$. Thus, unlike the conventional
scalar $QED_{2}$ of Eq. (\ref{a1}), this theory is infrared finite, at least 
to one loop order, in the vanishing mass limit. Consequently, it may be worth 
analyzing the properties of this model further.

First, let us note that the extra coupling in Eq. (\ref{a5}) is purely 
imaginary. This implies that the generalized theory in
Eq. (\ref{a5}), is  not Hermitian in Minkowski space.
However, it is worth noting that this theory is invariant under
the combined operations of PT. Such theories are of interest in recent 
literature \cite{bender}. Although non-Hermitian interactions may lead,
in general, to problems of unitarity, their effect in
this theory is not so problematic.
To understand this, we first remark that the model is actually
Hermitian in Euclidean space. It is well known \cite{das:book93},
that the path integral which defines the generating functional is
best evaluated by rotating to Euclidean space, where path integrals
are well behaved. Then, the massless scalar fields can be integrated
out in the path integral, leading to an exact action which is
Hermitian. After rotating back to Minkowski space, we obtain
\begin{equation}
\Gamma_{S}  =  -i \ln\left[{\det(-D_{\mu}D^{\mu}-{ie\over 2}\epsilon^{\mu\nu}
F_{\mu\nu}\sigma_{3})\over \det(-\partial_{\mu}\partial^{\mu})}
\right]^{-1} \label{a7}
\end{equation}
where $\Gamma_{S}$ represents the contribution of the scalar loops to the
effective action. Now, choosing
\begin{equation}
\gamma^{0} = \sigma_{2};\;\;\;\;\gamma^{1} = i\sigma_{1}; \;\;\;\;{\rm and}\;\;
\gamma_{5} = \gamma^{0}\gamma^{1} = \sigma_{3} \label{a8}
\end{equation}
and using the two-dimensional identity
\begin{equation}
\gamma^{\mu}\gamma^{\nu} = \eta^{\mu\nu} + \epsilon^{\mu\nu}\gamma_{5} 
\label{a9}
\end{equation}
it is easy to see that
\begin{equation}
\Gamma_{S}  =  -i \ln\left[{\det(-(D\!\!\!\!\slash)^{2})\over \det(-(\partial
\!\!\!\slash)^{2})}\right]^{-1}
  =  2i \ln\left[{\det(iD\!\!\!\!\slash)\over \det(i\partial\!\!\!\slash)}
\right]\label{a10}
\end{equation}
The quantity in the bracket, on the other hand, is the determinant for a 
massless fermion interacting minimally with the photon field in $1+1$ 
dimensions (Schwinger model). It is exactly soluble and is free from infrared 
problems leading
to\cite{schwinger:1962tp,das:1987yc}
\begin{equation}
\Gamma_{S} = -{e^{2}\over \pi} \int d^{2}x\;A_{\mu}\left(\eta^{\mu\nu} -
{\partial^{\mu}\partial^{\nu}\over \partial^{2}}\right)A_{\nu} \label{a11}
\end{equation}

Thus, we see that the generalized scalar $QED_{2}$ (Eq. (\ref{a5})) is 
exactly soluble in the massless limit and has the form similar to that 
obtained in the case of the massless Schwinger model. Although the 
additional interaction in (\ref{a5}) is non-Hermitian, the effective action 
for the theory is Hermitian,
\begin{equation}
{\cal L}_{eff} = -{1\over 4} F_{\mu\nu} F^{\mu\nu} - {e^{2}\over \pi}A_{\mu}
\left(\eta^{\mu\nu} - {\partial^{\mu}\partial^{\nu}\over \partial^{2}}\right)
A_{\nu} \label{a12}
\end{equation}
Even though the effective action in (\ref{a12}) has the same form as that
obtained in the case of the massless Schwinger model, there is one crucial 
difference.
Namely, the mass term for the photon has the wrong sign. We would like to 
emphasize that this is not a consequence of the additional non-Hermitian
interaction in Eq. (\ref{a5}). Rather, it is a reflection of the bosonic
nature of the fundamental fields being integrated out. (Fermions have an extra
negative sign coming from the loops, or equivalently the powers of the 
determinant for bosons is inverse of those for fermions\cite{das:book93}.) 
The additional 
interaction merely cancels the infrared divergence of the theory, as can be 
seen from (\ref{a3}) and (\ref{a6}). As a result of this, the effective
photon theory (\ref{a12}) is tachyonic and we can think of the tachyonic 
photon as a bound
state of scalar fields. Tachyonic quantum field theories have, of course,
been discussed in the literature\cite{feinberg:1967},
but we would like to emphasize that, unlike 
earlier discussions, this is a spin-1 theory of tachyons.

It is also interesting to note here that in a theory of scalars and fermions 
interacting with photons described by the Lagrangian density ($a=1,2$)
\begin{equation}
{\cal L} = -{1\over 4} F_{\mu\nu} F^{\mu\nu} + i\bar{\psi}_{a}D\!\!\!\!
\slash\psi_{a} + (D_{\mu}\phi_{a})^{*}D^{\mu}\phi_{a} - {ie\over 2}
\epsilon^{\mu\nu} F_{\mu\nu} \phi_{a}^{*}(\sigma_{3})_{ab}\phi_{b}
\label{a13},
\end{equation}
the radiative corrections due to bosons and fermions  cancel, leading to
an effective Lagrangian density
\begin{equation}
{\cal L}_{eff} = -{1\over 4} F_{\mu\nu} F^{\mu\nu}\label{a14}
\end{equation}
Namely, in such a theory, the photon would remain massless. This is quite 
interesting since radiative corrections due to fermions and bosons generally 
cancel in supersymmetric theories whereas the theory of Eq. (\ref{a13}) 
does not appear to have any supersymmetry. Indeed, we have not been able to 
find any conventional supersymmetry transformation which leaves the
theory of Eq. (\ref{a13}) invariant. 
Rather, we find several equivalent, BRST-like
nilpotent fermionic symmetries of this theory. For example, it is easy to
check that the theory of Eq. (\ref{a13}) is invariant under the symmetry 
transformations
\begin{eqnarray}
\delta\phi_{a} =  \sigma^+_{ab}\bar{\epsilon}
(1+\gamma_{5})\psi_{b} & {\;\;\; ;\;\;\;} &
\delta \phi_{a}^* =  \sigma^+_{ab}\bar{\psi}_b
(1+\gamma_{5})\epsilon
\nonumber \\
\delta \bar\psi_{a} =  -\sigma_{ab}^+\bar\epsilon(1+\gamma_{5})
\gamma^\mu(iD_\mu\phi_b)^* & {\;\;\; ;\;\;\;} &
\delta \psi_{a} =  -\sigma_{ab}^+
\gamma^\mu(iD_\mu\phi_b)(1+\gamma_5)\epsilon
\nonumber  \\
\delta A_\mu = 0
\end{eqnarray}
where $\sigma^+\equiv(1+\sigma_3)/2$ and
$\epsilon$ is a constant Majorana spinor. The fact that
these transformations do not form a self-adjoint set 
may be a reflection of the non-Hermitian nature of the 
Lagrangian density (\ref{a13}). 

Finally, we note 
that, instead of (\ref{a5}), if we consider a fermionic theory of the form
\begin{equation}
{\cal L}_{f} =  -{1\over 4} F_{\mu\nu} F^{\mu\nu} + (\overline{D_{\mu}\psi}) 
(D^{\mu}\psi) - {ie\over 2} \epsilon^{\mu\nu} F_{\mu\nu}\overline{\psi}\gamma_{5}
\psi \label{a16}  
\end{equation}
then, the additional interaction, in this case, would exactly represent the 
Pauli interaction and this Lagrangian density would be Hermitian. This is a
second order fermion Lagrangian density. Normally, higher derivative theories 
lead to problems of ghosts. However, in this case, it can be easily checked 
following the steps (\ref{a7})-(\ref{a10}) (with appropriate identifications)
that the effective action is identical to that obtained in the case of the 
massless Schwinger model. There is no problem of ghosts and that one can think
of Eq. (\ref{a16}) as a second order representation of the Schwinger model.

In conclusion, we have analyzed a generalized, massless scalar $QED_{2}$ and 
shown that it is free from infrared divergences. We have tried to bring out
various other interesting properties associated with this model.

A.D. is supported  in  part    by US  DOE  Grant number  DE-FG-02-91ER40685
and NSF-INT-9602559.  F.T.B and J.F. are partially
supported by CNPq (the National Research Council of Brazil) and 
F.T.B is supported in part by ``Programa de Apoio a N\'ucleos de
Excel\^encia'' (PRONEX).


\vfill

\begin{thebibliography}{1}

\bibitem{weinberg:book95}
See, for example,
S. Weinberg, {\em The Quantum Theory of Fields} (Cambridge University Press,
  Cambridge, England, 1995).


\bibitem{ferrara} S. Ferrara, Lett. Nuovo Cim. {\bf 13}, 629 (1975); A. 
Smailagic and J. A. Helayel-Neto, Mod. Phys. Lett. {\bf A2}, 787 (1987); R. 
Amorim and A. Das, phys. Rev. {\bf D57}, 2599 (1998).

\bibitem{bender} See, for example, C. Bender, S. Boettcher, P. Meisinger, 
quant-ph/9809072, and references therein.

\bibitem{das:book93}
A. Das, {\em Field Theory: A Path Integral Approach} (World Scientific, NY,
  1993).

\bibitem{schwinger:1962tp}
J. Schwinger, Phys. Rev. {\bf 128},  2425  (1962).

\bibitem{das:1987yc}
See also A. Das and V. S. Mathur, Phys. Rev. {\bf D33},  489  (1986) and 
references therein.

\bibitem{feinberg:1967}
G. Feinberg, Phys. Rev. {\bf 159},  1089  (1967).

\end{thebibliography}
\end{document}